# Mapping groundwater potential zones in Cilongok Area, Banyumas, Central Java using 2D geoelectrical resistivity

*Sandy Hardian Susanto Herho* [1], *Plato Martuani Siregar* [2], *Iqbal Syarief* [3], *Dasapta Erwin Irawan* [4], *Marlison Justiana Sinaga* [5]

[1,2] Atmospheric Science Research Group, Faculty of Earth Sciences and Technology, Bandung Institute of Technology

[3,4] Applied Geology Research Group, Faculty of Earth Sciences and Technology, Bandung Institute of Technology

[5] PT Tirta Anugrah Geoservindo (TAG), Jakarta

E-mail: sandyherho@meteo.itb.ac.id

**Abstract**

The aim of this study is to investigate the presence of potential aquifer layers that would be used as a source of clean water for residents in Cilongok sub-district. We conducted a ground water potential survey at two locations in the sub-district, i.e. Cikidang and Kalisari villages. At each location, geoelectric measurements were taken with the Wenner configuration and the horizontal distance of 900 meters, so that we can get the estimation of lithology layers that potentially store groundwater up to the depth of ± 100 meters. Then the measurement result was processed by using Res2DInv software to get the 2D resistivity images of rock layers. Based on the resistivity estimation and observation of regional geological conditions in that area, then we conducted 2D interpretation about potential lithology layers that economically potential for water well drilling and to be utilized as a source of clean water for the residents. Based on the interpretation of lithology, it was known that the aquifer layer that has the potential as a source of clean water is a layer that consists of coarse and porous tuffs from volcanic eruption products that can easily drain groundwater on the average about 28 to 45 meters depth in Cikidang village and about 45 to 60 meters depth in Kalisari village.

**Introduction**

Cilongok sub-district became one of the areas affected by environmental pollution due to land clearing and groundwater exploitation due to PT Sejahtera Alam Energi's operational activities in Mount Slamet (SatelitPost, 2017). Whereas Cilongok sub-district is one of the centres of tofu home industry in Banyumas Regency, but Cilongok Sub-district has clean water demand in some areas, such as Karangtengah, Panembangan, Kalisari, Karanglo, and Cikidang (Radar Banyumas, 2017). This is evident from the lack of water sanitation in the villagers settlements that even disrupt their sanitation activities. Therefore, groundwater exploration activities are needed to investigate the availability of clean water in the villages.

Surveying activities through the soil or underground should be done to obtain an arrangement of layers of the earth, to be known the presence of water carriers (aquifer), thickness and depth and to take water samples for water quality analysis. Although groundwater cannot be directly observed through the Earth's surface, ground-level investigations are a preliminary investigation, which at least can provide an idea of the location of the groundwater presence. Several ground-level methods of inquiry can be undertaken, including geological methods, gravity methods, magnetic methods, seismic methods, and geoelectric methods. Of these methods, the geoelectric method is a widely used method and the result is quite good (Bisri, 1991).

This geoelectric estimation is based on the fact that different materials will have different types of resistance when electrified. The geoelectrical method is done by measuring the potential difference caused by injection of electric current into the earth. The properties of a formation can be illustrated by three basic parameters: electrical conductivity, magnetic permeability, and dielectric permittivity (Siregar, 2014). The conductivity properties of porous rocks are generated by the conductivity properties of fluid filling the pores, pore space interconnections and the conductivity properties of granular interfaces and pore fluids (Reynolds, 2011). Based on its electrical resistivity value, a subsurface structure can be known by its constituent material. The geoelectrical method is quite simple, cheap and very susceptible to interference so it is suitable for use in shallow groundwater exploration. Design of monitoring system using electrical resistivity is very important to detect groundwater flow (Siregar, 2014). Through this geoelectric survey method, it is expected to find the location of the abundance of clean groundwater that can be accessed by residents of Kalisari and Cikidang villages directly.

**Regional Geology**

The location of this study is in Banyumas Regency covering Kalisari and Cikidang villages which are intended to be around coordinates 291193.5 E/ 9183923.1 S/ 49 S and 293361.7 S/ 9181809.2 S/ 49 S. Based on Geological Map of Purwokerto and Tegal quadrangles (Djuri et al., 1996), the Kalisari and Cikidang villages are included in the North Serayu Mountainious Zone. Based on the stratigraphical order, Kalisari and Cikidang areas are classified as Holocene young rock formations, namely

Slamet lahar deposits that consist of andesite-basaltic rocks that are 10 - 50 cm in diameter. Slamet lahar deposits was a product of the Old Slamet volcanic eruptions. Based on regional tectonic setting, Cilongok sub-district is located between North and South Serayu mountainous zone, specifically in the Intermountain region, where there are four upthrust and some normal faults, these faults are estimated to occur at tectonic events occurring around Miocene - Pliocene which is accompanied by the formation of the intrusive rock (Suranto, 2008). Based on the rock units pattern, Kalisari and Cikidang areas consist of two rock units, i.e. the Andesite unit and the Volcanic breccia unit (Kusmayadi et al., 2014) (Figure 1).

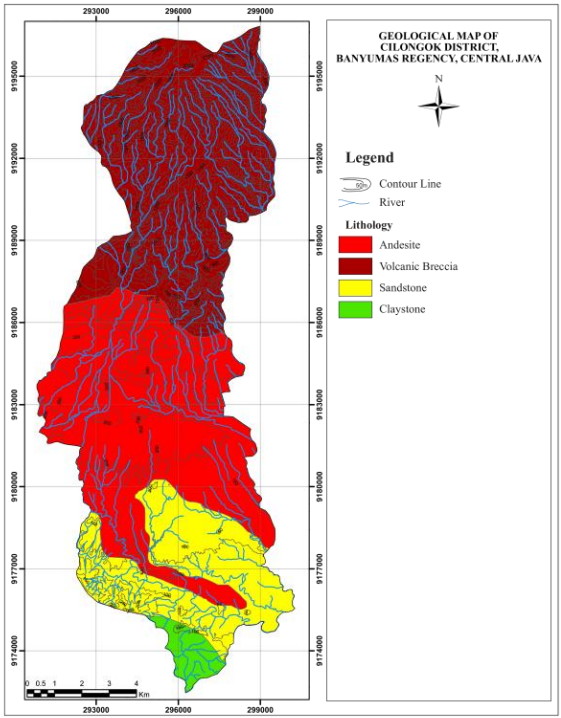

Figure 1: Lithology Map of Cilongok sub-district. (modification from Kusmayadi et al., 2014).

Therefore the Kalisari and Cikidang areas are located on the hillside of Slamet Volcano so that the local hydrology conditions are affected by hydrogeological properties of the Slamet lahar deposits. Based on the stratovolcano area hydrology characteristics, as well as the geological condition of the local volcanic rock units, the aquifer conditions of the study area fall into the category of local productive aquifers. The aquifer layers in the Slamet Volcano hillside is aquifer with water flowing through the fissures and inter-grain spaces. The aquifers in this area are locally productive with very deep groundwater level, and there are some small discharge springs that can be utilized by the people (Suranto, 2008).

## Background Theory

The geoelectrical resistivity method is based on the assumption that the earth has a homogeneous isotropic property. With this assumption, the measured resistivity is a true resistivity and does not depend on the electrodes spacing. But in reality the earth is made up of layers with different resistivity, so the measured electrical voltage was the influence of the layers ($\rho_a$). This apparent resistivity value can be determined by applying equation (1).

$$\rho_a = \frac{2\pi}{\left[\left(\frac{1}{r_1}-\frac{1}{r_2}\right)-\left(\frac{1}{r_3}-\frac{1}{r_4}\right)\right]} \cdot \frac{\Delta V}{I} \qquad (1)$$

Or the apparent resistivity value can also be calculated by equation (2).

$$\rho_a = K\frac{\Delta V}{I} \qquad (2)$$

The geometry $K$ factor is the magnitude of the correction of the location of the two potential electrodes to the location of the current electrode (Reynolds, 2011).

In this study, we used the Wenner configuration in the geoelectrical survey. The Wenner configuration is one of the most commonly used configurations in geoelectrical survey (Reynolds, 2011), with the same long spaces spacing ($r_1 = r_4 = a$ and $r_2 = r_3 = 2a$). The distance between the current electrode is three times the distance of the potential electrode, the potential distance with the sounding point is $a/2$, the distance of each current electrode with its sounding point is $3a/2$. The target depth that can be achieved by this method is $a/2$. In the field data acquisition, the current and potential electrode arrangement is placed symmetry with the sounding point.

In the Wenner configuration, the distance between the current electrode and the potential electrode is the same, as shown in Figure 2. In Figure 2, it shows that the distance $C_1P_1 = C_2P_2 = a$ and the distance $C_1P_2 = C_2P_1 = 2a$, then we can calculate the geometry factor of the Wenner configuration and the apparent resistivity of a layer through equations (3), (4), and (5).

$$K = \frac{2\pi}{\left[\left(\frac{1}{a}-\frac{1}{2a}\right)-\left(\frac{1}{2a}-\frac{1}{a}\right)\right]} \qquad (3)$$

, therefore:
$$K = 2\pi a \qquad (4)$$

, and:
$$\rho_a = K.R \qquad (5)$$

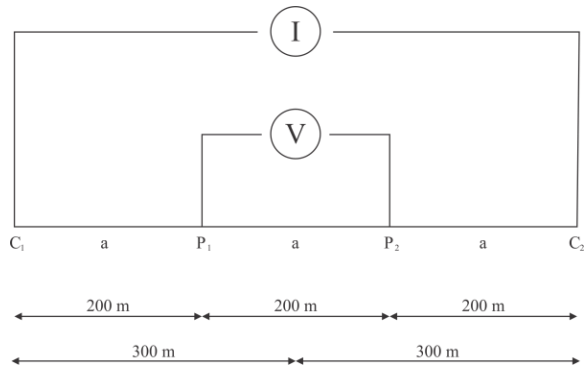

Figure 2: Current and potential electrodes in Wenner configuration.
(modification from Dobrin and Savit, 1988).

**Data and Methods**

To obtain the depth data and the existence of the aquifer layers, the steps we undertook were to study the geology around the study area; to determine the measurement points with a space of 200 m using Wenner's configuration (to obtain a depth of ± 100 m, the expansion of the cable should reach $(C_1 - C_2)/2 = 300$ m as shown in Figure 2); selecting a sloping lines data path < 20 ° (Dobrin and Savit, 1988); performs 2D rock layers resistivity data plot by using Res2DInv software (Loke, 2002) for each observational lines, and finally to interpret the geological layer of rocks based on resistivity image of rock layers in both study locations, i.e. Cikidang and Kalisari villages. We conducted the geoelectrical survey on twelve trajectory lines with details of six lines in the Cikidang area and six lines in the Kalisari region (Figure 3). The types of equipment that we used when conducting this geoelectrical survey are Nainura NRD 300 resistivity meter; the electrodes; cable roll; NLG EC1500AW Generator Set 1KVA; Garmin Handheld GPS; hammer; meter; laptop; stationeries; camera; and the geological map of the survey location (Figure 4). We interpreted the resistivity value of each layer by matching with the table of rock resistivity values that have been interpreted by Helide (1984) (Table 1).

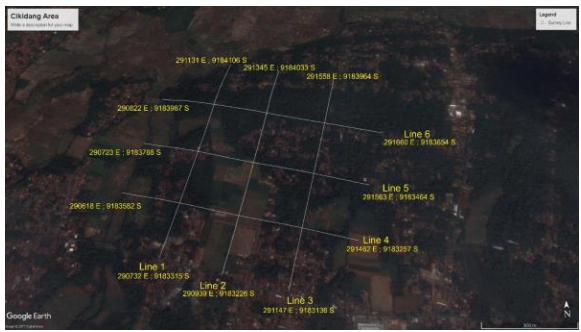

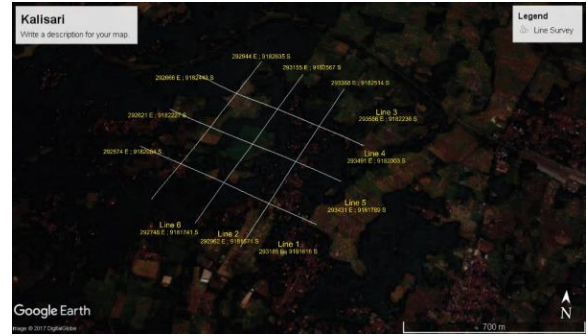

Figure 3: Location of geoelectrical survey in regions (a) Cikidang and (b) Kalisari.

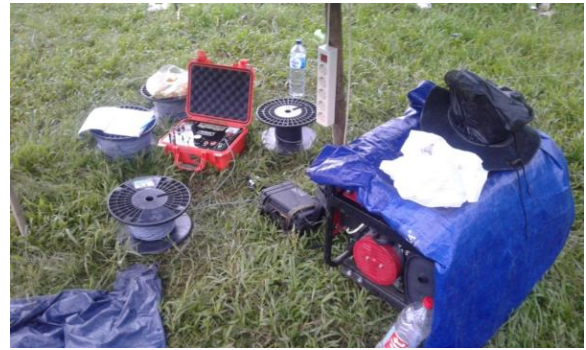

Figure 4: Types of equipment that used for geoelectrical measurement.

Table 1: List of the resistivity value of some *subsurface* rock materials.
(Helide, 1984)

| Material Types | Resistivity value (Ωm) |
|---|---|
| Surface water | 80 – 200 |
| Groundwater | 30 – 100 |
| Water in alluvial layers | 10 – 30 |
| Spring water | 50 – 100 |
| Sand and gravel | 100 – 1000 |
| Sand and gravel that containing fresh water | 50 – 500 |
| Sand and gravel that containing salt water | 0.5 – 5 |
| Mudstone | 20 – 200 |
| Conglomerate | 100 – 500 |
| Clay | 2 – 20 |
| Marl | 20 -200 |
| Tuff | 0,5 – 5 |
| Limestone | 300 – 10000 |
| Lava | 300 – 10000 |
| Shale that containing granite | 0.5 – 5 |
| Shale interbedded with clay | 100 – 300 |
| Clay sandstone | 50 – 300 |
| Quartzite sandstone | 300 – 10000 |
| Gneiss with some granite | 100 – 1000 |
| Granite | 1000 – 10000 |

**Results and Discussions**

The following paragraphs are the interpretation of the existence of the suspected aquifer layers in the Cikidang observation area:

In the location of **Cikidang Line 1**, there is an alluvial layer which has a resistivity value <20 Ωm at a depth of 0 - 28 meters. Then there is a coarse tuffs layer at a depth of about 28 - 75 meters marked by a dark green coating shown in Figure 5 which allegedly stores groundwater with a resistivity value of 20 - 60 Ωm. Then at depths below 75 meters, there are suspected of Slamet Volcano lava deposits.

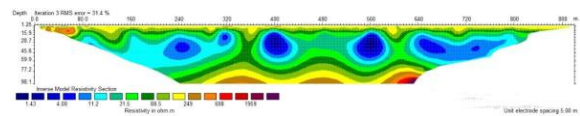
Figure 5: 2D Resistivity Profile in **Cikidang Line 1**.

In the location of **Cikidang Line 2** (Figure 6), the layer at a depth of 0-16 meters is alluvial soil with a resistivity of <20 Ωm. Layers at depths of 16 - 45 meters are thought to be coarse tuffs with 20 - 60 Ωm resistivity containing freshwater. Layers below 45 meters are strongly suspected as a layer of volcanic lava that does not contain groundwater.

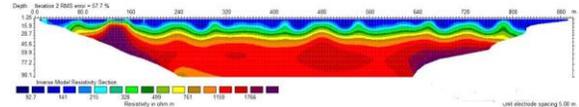
Figure 6: 2D Resistivity Profile in **Cikidang Line 2**.

The geoelectric survey in **Cikidang Line 3** (Figure 7) shows that the layer at a depth of 0-16 meters is alluvial soil with a resistivity of <20 Ωm. Layers at depths of 16 - 50 meters are thought to be coarse tuffs with 20 - 60 Ωm resistivity containing freshwater. Layers below 50 meters are strongly suspected as a layer of volcanic lava that does not contain groundwater.

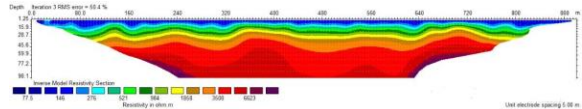
Figure 7: 2D Resistivity Profile in **Cikidang Line 3**.

The geoelectric survey of **Cikidang Line 4** (Figure 8) shows that the layer at a depth of 0 - 15 meters is alluvial soil with a resistivity value <20 Ωm. Layers at a depth of 15 - 45 meters are thought to be coarse tuffs with 20 - 60 Ωm resistivity containing freshwater. Layers below 45 meters are strongly suspected as a layer of volcanic lava that does not contain groundwater.

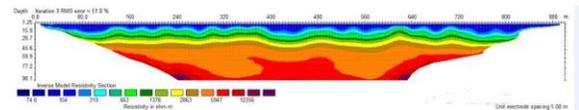
Figure 8: 2D Resistivity Profile in **Cikidang Line 4**.

Geoelectric survey results in **Cikidang Line 5** (Figure 9), shows that at an average depth of 0-20 meters the layer is composed of alluvial layers with a resistivity value <20 Ωm. At a depth of 20 - 60 meters potentially found a coarse tuff layer with a resistivity value of 20 - 60 Ωm containing fresh water. Below the depth of 60 meters there is only Slamet volcano lava deposits that does not contain groundwater.

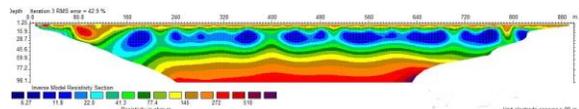
Figure 9: 2D Resistivity Profile in **Cikidang Line 5**.

Geoelectric survey results in **Cikidang Line 6** (Figure 10) shows that at a depth of 0 - 5 meters there is water in alluvial soil layers having a resistivity value <20 Ωm. The next layer is a coarse tuffs layer which has a resistivity value of 20 - 60 Ωm at a depth of 3 - 30 meters.

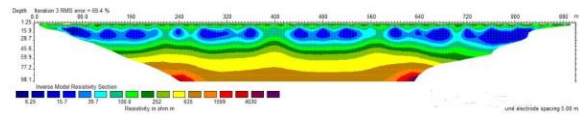

Figure 10: 2D Resistivity Profile in **Cikidang Line 6**.

If groundwater exploration is based on a coarse tuff aquifer of 20 - 60 Ωm, then an interpretation of the geoelectric measurement results is as follows. In **Cikidang Line 1**, groundwater is likely to be found at a depth of 28 - 75 meters. In the **Cikidang Line 2** site, groundwater is likely to be found at a depth of 16 - 45 meters. In **Cikidang Line 3**, groundwater is likely to be at a depth of 16 - 50 meters. In the location of **Cikidang Line 4**, it is possible to find groundwater at a depth of 15 - 45 meters. In the location of **Cikidang Line 5**, groundwater is in position 20 - 60 meters below the surface. In **Cikidang Line 6**, groundwater is likely to be found, at a depth of 5 - 30 meters.

Based on the results of the lithologic assessment of geoelectrical and regional geological interpretations of Cikidang observational area, then we do the interpretation in the form of lithology fence diagram which can be seen in Figure 11.

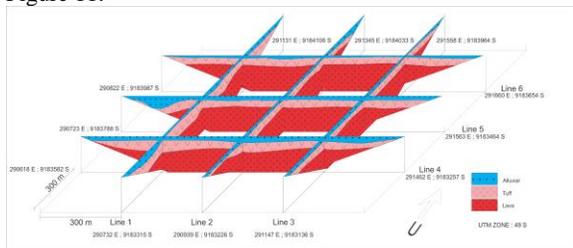

Figure 11: Estimated fence diagram of Cikidang area lithology.

The following paragraphs are the interpretation of the existence of the suspected aquifer layers in the Kalisari observation area:

The geoelectric survey of **Kalisari Line 1** (Figure 12) shows that the layer at 0 - 30 meters depth on Kalisari Line 1 is an alluvial soil layer that looks yellow in Figure 14 with a resistivity value <20 Ωm. At a depth of 30 - 70 meters there is a coarse tuffs layer marked in green with the cavities marked in blue, potentially as aquifer (Figure 12) with a resistivity value of 20 - 60 Ωm. At depths below 70 meters, there are Slamet Volcano lava deposits, so it is impossible to find groundwater in the layer.

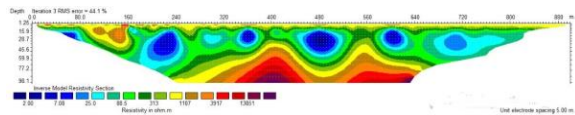
Figure 12: 2D Resistivity Profile in **Kalisari Line 1**.

The geoelectric survey in **Kalisari Line 2** (Figure 13) shows that the layer at depths of 0-27 meters is an alluvial soil layer with a resistivity value <20 Ωm marked with dark blue in Figure 13. Then a layer of light blue to dark green colours at a depth of 27 - 45 meters is a coarse tuffs layer containing fresh water with a resistivity value of 20 - 60 Ωm. While layers below 45 meters allegedly Slamet Volcano lava deposits and other bedrocks that have a very high resistivity value, were difficult to find groundwater.

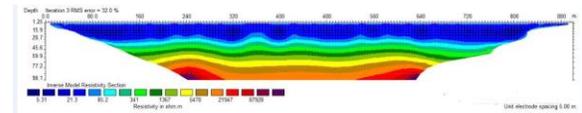
Figure 13: 2D Resistivity Profile in **Kalisari Line 2**.

The geoelectric survey of **Kalisari Line 3** (Figure 14) shows that the first layer at 0-50 meters depth is an alluvial soil layer with a resistivity value <20 Ωm which is indicated by light blue in Figure 14. In the next layer, it is assumed that there is a coarse tuff containing freshwater at a depth of 50 - 77 meters marked from dark blue to beige in Figure 14 with a resistivity value of 20 - 60 Ωm. Layers below 77 meters allegedly Slamet Volcano lava deposits is difficult to find groundwater in it.

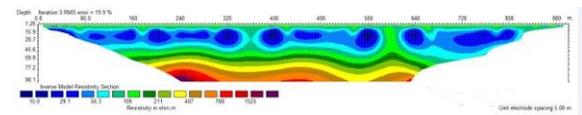
Figure 14: 2D Resistivity Profile in **Kalisari Line 3**.

The geoelectric survey in **Kalisari Line 4** (Figure 15) shows that at a depth of 0 - 40 meters there is an alluvial layer with a resistivity value <20 Ωm. Between 40 - 77 meters depth there is a layer of coarse tuffs that is suspected to contain fresh water with a resistivity value between 20 - 60 Ωm. Below the depth of 77 meters, it is difficult to find groundwater in layers with a resistivity of more than 500 Ωm.

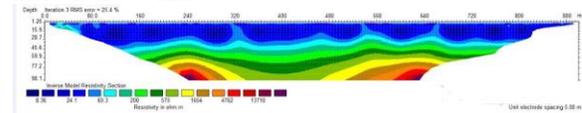
Figure 15: 2D Resistivity Profile in **Kalisari Line 4**.

The geoelectric survey results in **Kalisari Line 5** (Figure 16), shows that on the surface to a depth of 45 meters there are alluvial soil layers with resistivity values between <20 Ωm. At a depth of 45 - 60 meters it is suspected that there is a coarse tuff layer containing fresh water with a resistivity value of 20 - 60 Ωm. The underlying layer is thought to be the Slamet Volcano lava deposits, making it difficult to find groundwater.

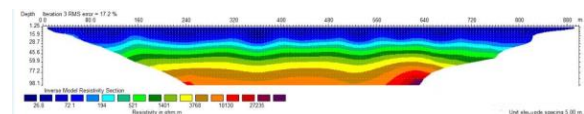
Figure 16: 2D Resistivity Profile in **Kalisari Line 3**.

Particularly in the **Kalisari Line 6** observation location, there are two interpretations:
- **Interpretation based on geoelectrical data** (Figure 17): The layer at a depth of 0 - 10 meters is an alluvial soil layer that does not contain fresh water with a resistivity value of <20 Ωm which is indicated by light blue in Figure 17. Then in the next layer to about 60 meters depth there is a Slamet Volcano lava intrusion that also does not contain fresh water. Assuming a coarse tuff layer containing fresh water has a resistivity value of 20 - 60 Ωm, then the possibility of fresh water on **Kalisari Line 6** can be found at depths of more than 70 meters.

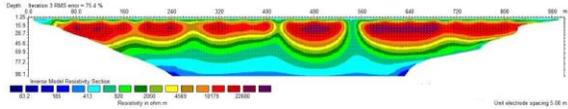

Figure 17: 2D Resistivity Profile in **Kalisari Line 6**.

- **Interpretation based on regional geological conditions** (Figure 18): Alluvial soil layer at a depth of 0 - 30 meters, the second layer is a coarse tuff layer with a depth of 30 - 75 meters, then the bottom layer is Slamet Volcano lava deposits at depths below 75 meters.

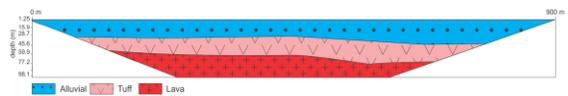

Figure 18: Lithology cross-section in **Kalisari Line 6**.

If groundwater exploration is based on a coarse tuff aquifer of 20 - 60 Ωm, then an interpretation of the geoelectric measurement results is as follows. In **Kalisari Line 1**, groundwater is likely to be found at a depth of 30 - 70 meters. In the **Kalisari Line 2** location, groundwater is likely to be found at a depth of 27 - 45 meters. In **Kalisari Line 3**, groundwater is likely to be at a depth of 50 - 77 meters. In the **Kalisari Line 4** location, it is possible to find groundwater at a depth of 40 - 77 meters. In the **Kalisari Line 5** location, groundwater is in the position of 45 - 60 meters below the surface. In **Kalisari Line 6**, groundwater is likely to occupy a depth below 70 meters (according to resistivity estimation), whereas according to geological interpretation at 30 - 75 meters depth.

Based on the results of the lithologic assessment of geoelectrical and regional geological interpretations of Kalisari observational area, then we do the interpretation in the form of lithology fence diagram which can be seen in Figure 19.

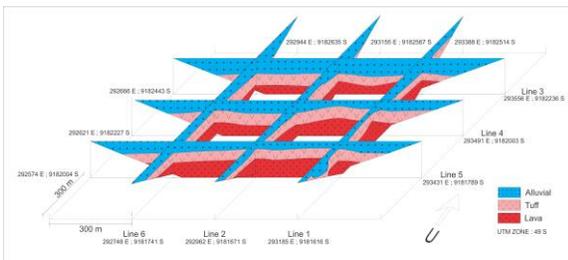

Figure 19: Estimated fence diagram of Kalisari area lithology.

## Conclusion

Based on the results of the geoelectrical at various locations in Kalisari and Cikidang villages and regional geological interpretations to both villages, there is some potential groundwater source below the surface at an average depth of 28-45 meters (Cikidang), and 45 - 60 meters (Kalisari). Kalisari and Cikidang villages located on the hillside of Slamet Volcano which has relatively flat topography with the surface rock in the form of Slamet volcanic eruption products. Coarse tuffs from the volcanic eruption products have many spaces and can easily distribute groundwater. Considering only the thickness of the coarse tuffs aquifer layer and its proximity to the surface (technical aspect) and without considering the legal aspects (land ownership), the most potential location for exploration of groundwater drilling at the Kalisari survey location is at coordinates, 293185 E / 9182003 S / 49 S, while at the Cikidang survey location at 290732 E / 918788 S / 49 S coordinates.

**Acknowledgements**

This research was partially supported by Weather and Climate Prediction Laboratory (WCPL) ITB. We acknowledge the management of PT Sejahtera Alam Energi (SAE) for their support and their permission to publish and present this paper. We also thank Abdul Haris Wirabrata for comments that greatly improved the manuscript.